\documentclass[12pt,draftcls,onecolumn]{article}

\usepackage{nidanfloat}
\usepackage{flushend}
\usepackage{multirow}
\usepackage{amssymb, bm, setspace, comment}
\usepackage{cite}
\usepackage[pdftex]{graphicx}
\usepackage{epsfig}
\usepackage[cmex10]{amsmath}
\usepackage[skip=1pt]{subcaption}
\usepackage{amssymb}
\usepackage{bm}

\graphicspath{{./images/}}

\begin{document}

\title{Complex-valued reservoir computing for aspect classification and slope-angle estimation with low computational cost and high resolution in interferometric synthetic aperture radar}

\author{Bungo~Konishi,~Akira~Hirose,~and~Ryo~Natsuaki}

\date{}

\maketitle

\begin{abstract}
Synthetic aperture radar (SAR) is widely used for ground surface classification since it utilizes information on vegetation and soil unavailable in optical observation.
Image classification often employs convolutional neural networks.
However, they have serious problems such as long learning time and resolution degradation in their convolution and pooling processes.
In this paper, we propose complex-valued reservoir computing (CVRC) to deal with complex-valued images in interferometric SAR (InSAR).
We classify InSAR image data by using CVRC successfully with a higher resolution and a lower computational cost, i.e., one-hundredth learning time and one-fifth classification time, than convolutional neural networks.
We also conduct experiments on slope angle estimation.
CVRC is found applicable to quantitative tasks dealing with continuous values as well as discrete classification tasks with a higher accuracy.
\end{abstract}

\renewcommand{\thefootnote}{}

\footnote{
A part of this work was supported by JSPS KAKENHI under Grant 18H04105 and also by Cooperative Research Project Program of the Research Institute of Electrical Communication (RIEC), Tohoku University. The Advanced Land Observing Satellite (ALOS) original data are copyrighted by the Japan Aerospace Exploration Agency (JAXA) and provided under JAXA Fourth ALOS Research Announcement PI No. 1154 (AH).
}%
\footnote{
The authors are with the Department of Electrical Engineering and Information Systems, The University of Tokyo, Tokyo 113-8656, Japan (e-mail: b.konishi@eis.t.u-tokyo.ac.jp; ahirose@ee.t.u-tokyo.ac.jp; natsuaki@ee.t.u-tokyo.ac.jp).
}%
\footnote{
A part of this work was presented in Three Minute Thesis (3MT) Competition in IGARSS 2020 and received 3rd prize.
}%

\renewcommand{\thefootnote}{\arabic{footnote}}

\section{Introduction}
\label{s:introduction}

Synthetic aperture radar (SAR) is an active sensor to enhance the spatial resolution by synthesizing a large aperture by moving a platform such as an airplane or satellite.
On the other hand, many studies have reported that hyperspectral optical sensors are utilized in estimation of vegetations and soils \cite{Cheng2018a,Cheng2018b,Marmanis2016}.
However, optical sensors have a problem of its unavailability under some climate and time conditions.
SAR data includes information on phase and polarization that optical sensors cannot obtain.
Hence, an appropriate processing of such information will lead to better classification of land surfaces \cite{Ulaby2015}.

The topographical information includes slope, aspect, and elevation.
Among others, aspect and slope are widely utilized for land use planning, hazard mapping and natural gus-pipeline routing because such features provide the most important data for assesment of landslide susceptibility \cite{Marsala2019,Cevik2003}. 

Deep neural networks adaptively classify land surfaces with high accuracy in many recent studies \cite{De2018,Lin2017,Zhou2016,Geng2017,Ding2016}.
Since a SAR image is composed of complex values, it is required to be processed by using complex-valued neural networks (CVNNs) \cite{Hirose2012a}.
Suksmono and Hirose demonstrated that CVNNs have an advantage over real valued neural networks (RVNNs) in land form segmentation in interferometric SAR (InSAR) data \cite{Suksmono2000}. This is because CVNNs deal with both amplitude and phase information with high generalization ability \cite{Hirose2012b} whereas RVNNs only deal with amplitude information or real and imaginary part separately.
In addition, they utilized CVNNs as a method to remove phase singular points \cite{Suksmono2002a,Suksmono2002b}.

Zhang proposed complex-valued convolutional neural networks (CVCNNs) by combining convolutional neural networks (CNNs) with CVNNs to classify land use such as vegetation and soil in polarimetric SAR data \cite{Zhang2017}.
As the results, they demonstrated that CVCNNs have a higher accuracy than RVNNs.
On the other hand, Sunaga proposed another CVCNNs in order to classify and discover land forms \cite{Sunaga2019}. 
Thereby, they demonstrated that CVCNNs can extract land forms well which have similar shapes.

CNN is generally consisting of alternating convolution and pooling layers.
The pooling process takes the averages or maximum value in a local window to absorb dislocations which are harmful in recognition tasks. 
This is one of the strengths in image recognition, while it causes resolution degradation in image segmentation.

In addition, these methods using neural networks require high computational cost because they search an approximate solution by iterative updating of network weights by using a gradient algorithm.
However, SAR observes the ground with various polarization, frequency and/or spatial resolution. 
Then, the total computational cost becomes very high in the learning for various platforms and diverse analyses.

Reservoir computing (RC) \cite{Jaeger2001,Maass2002}, a neural network-based framework, has big potential for a high speed learning without resolution degradation \cite{Tanaka2019}.
A reservoir has multiple neurons recurrently connected and maps input signals to a high dimensional space.
The recurrent connection weights are fixed, while only connection weights between the reservoir and an output layer are trained.
This configuration contributes to an expeditious learning.
Besides temporal information processing, RC is used for image recognition with conversion of spatial data into serial data.

Recently, several studies have reported that RC has a high performance compared with CNNs on classification tasks of handwritten character images.
Besides, Jalalvand et al. demonstrated that RC has robustness against images including various noises \cite{Jalalvand2016,Jalalvand2015,Jalalvand2018}.
Additionally, Tong and Tanaka demonstrated that RC has a high classification performance in general without heuristic preprocessing \cite{Tong2018}.

Since SAR image is handled in the complex domain, the method is required to have a generalization ability in the complex domain, which has required a high computational cost.
Therefore, in this paper, we propose complex-valued RC (CVRC) to deal with complex-valued data of InSAR in low computational cost.

We classify land forms and estimate slope angles by using CVRC to process InSAR complex-amplitude data obtained by Advanced Land Observing Satellite (ALOS) of Japan Aerospace Exploration Agency (JAXA).
Experiments of aspect classification demonstrate that CVRC classifies local land forms more accurately, showing a high generalization ability in the complex domain, with about one-hundredth learning time and one-fifth classification processing time compared with CVCNN.

The rest of the paper is organized as follows.
Section \ref{s:proposal} describes the dynamics of CVRC we propose. 
In Section \ref{s:aspect}, we explain our experimental setup for aspect classification and present the results.
We also discuss the features specific to CVRC in comparison with other methods.
In Section \ref{s:slope}, we demonstrate that CVRC achieves higher performance in slope angle estimation.
Section \ref{s:conclusion} concludes this paper.

\section{Proposal of complex-valued reservoir computing}
\label{s:proposal}

\subsection{Dynamics of complex-valued reservoir computing}

\begin{figure}[tp]
  \centering
  \includegraphics[width=0.98\linewidth]{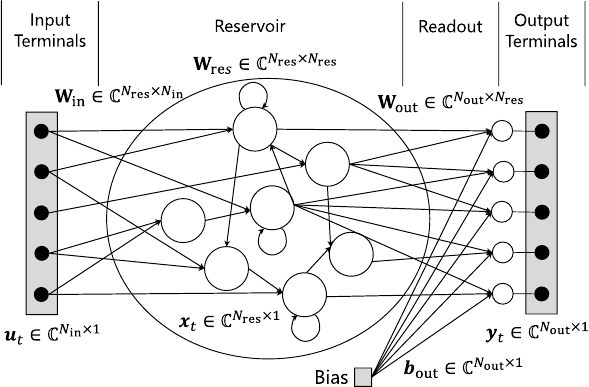}
  \caption{Structure of Complex-valued reservoir computing}
  \label{fig:cvrc}
\end{figure}

In this section, we propose CVRC to deal with complex values.
Fig.~\ref{fig:cvrc} shows the structure of the CVRC network.
The network has $N_\mathrm{in}$ input terminals in the input layer, $N_\mathrm{res}$ neurons in the reservoir and $N_\mathrm{out}$ neurons in the output layer.
The vectors $\bm{u}_t$, $\bm{x}_t$ and $\bm{y}_t$ represent input signals, output signals of neurons in the reservoir and output signals of those in the output layer, respectively.
The $N_\mathrm{res}\times N_\mathrm{in}$ matrix $\mathbf{W}_\mathrm{in}$ is the weights connecting the input layer to the reservoir, the $N_\mathrm{res}\times N_\mathrm{res}$ matrix $\mathbf{W}_\mathrm{res}$ is the recurrent connection weights and the $N_\mathrm{out}\times N_\mathrm{res}$ matrix $\mathbf{W}_\mathrm{out}$ is the weights connecting the reservoir to the output layer.
The vector $\bm{b}_\mathrm{out}$ is the bias of the output layer.
All of the signals, weights and bias have complex values.

The forward processing of the network is expressed as
\begin{align}
  \label{formula:cvrc_z}
  \bm{z}_t &= \mathbf{W}_\mathrm{in} \bm{u}_t+\mathbf{W}_\mathrm{res} \bm{x}_{t-1}\\
  \label{formula:cvrc_x}
  \bm{x}_t &= \left(1-\alpha \frac{\delta}{c}\right) \bm{x}_{t-1}+\frac{\delta}{c} \tanh(|\bm{z}_t|)\circ \exp(j \arg(\bm{z}_t))\\
  \label{formula:cvrc_y}
  \bm{y}_t &= \mathbf{W}_\mathrm{out} \bm{x}_t + \bm{b}_\mathrm{out}
\end{align}
where $\tanh(\cdot)$ is a hyperbolic tangent function, $|\cdot|$ is an amplitude and $\arg(\cdot)$ is a phase.
These operations are applied element-wise.
$\delta > 0$ is a discritization stepsize, $c > 0$ is a global time constant, $\circ$ is the Hadamard product and $\bm{z}_t$ denotes the internal states of neurons.
A leaking rate $\alpha$ ($0\leq\alpha\leq 1$) represents a hyperparameter introduced in order to determine the dynamics speed.
When $\alpha$ is small, the network retains past information for a long time in the reservoir.
Contrarily, when $\alpha$ is large, the reservoir forgets past information in a short time.

The dynamics (\ref{formula:cvrc_x}) is inconvenient for application use because we need to adjust many hyperparameters.
Therefore, we applied the following simplified dynamics, instead of (\ref{formula:cvrc_x}) by following \cite{Lukovsevivcius2012}, expressed as
\begin{align}
  \label{formula:cvrc_x2}
  \bm{x}_t &= (1-\alpha) \bm{x}_{t-1}+\alpha \tanh(|\bm{z}_t|)\circ \exp(j \arg(\bm{z}_t))
\end{align}
The CVRC can also describe continuous-time dynamics by giving
\begin{align}
  \label{formula:cvrc_z_c}
  \bm{z} &= \mathbf{W}_\mathrm{in} \bm{u}+\mathbf{W}_\mathrm{res} \bm{x}\\
  \label{formula:cvrc_x_c}
  \frac{d\bm{x}}{dt} &= \frac{1}{c}  \left(-\alpha \bm{x}+\tanh(|\bm{z}|)\circ \exp(j \arg(\bm{z})) \right)\\
  \bm{y} &= \mathbf{W}_\mathrm{out} \bm{x} + \bm{b}_\mathrm{out}
\end{align}

The CVRC differs from conventional RVRC in its activation function of the reservoir neurons.
In order to deal with waves, we define the activation function with saturated amplitude and retained phase so that the nonlinearity of the amplitude represents energy saturation and the phase corresponds to the rotation.
Therefore, the signal transformation becomes independent of phase reference in the SAR measurement, namely, the real and imaginary axes that determines in-phase and quadrature components.


A spectral radius $\sigma(\mathbf{W}_\mathrm{res})$, the absolute maximum eigenvalue of $\mathbf{W}_\mathrm{res}$, is also one of the important indications deciding remaining time.
Generally, $\sigma(\mathbf{W}_\mathrm{res})$ below unity is employed for maintaining stability of the system.
The following normalization treatment enables the network to adjust $\mathbf{W}_\mathrm{res}$ and decides a desirable spectral radius $\sigma_\mathrm{d}$. 
\begin{equation}
  \mathbf{W}_\mathrm{res} \leftarrow \frac{\sigma_\mathrm{d}}{\sigma(\mathbf{W}_\mathrm{res})}\mathbf{W}_\mathrm{res}
\end{equation}

We calculate the optimal $\mathbf{W}_\mathrm{out}$ and $\bm{b}_\mathrm{out}$ by using Tikhonov regularization based on the linear regression in (\ref{formula:cvrc_y}) as
\begin{equation}
  \label{formula:learning}
  \begin{bmatrix}
    \mathbf{W}_\mathrm{out} & \bm{b}_\mathrm{out}
  \end{bmatrix}
  = ((\mathbf{X}^*  \mathbf{X} + \lambda \mathbf{I})^{-1}  \mathbf{X}^*  \mathbf{D})^T
\end{equation}
where $\lambda$ is a regularization parameter and $\mathbf{X}$ ($\in \mathbb{C}^{N\times N_\mathrm{res}+1}$) is the matrix of aligned $\bm{x}_{t_i} (i=0,\cdots,N)$ corresponding to all teacher signals as follows.
We append one to $\mathbf{X}$ at each row to calculate $\bm{b}_\mathrm{out}$ as
\begin{equation}
  \label{formula:x-def}
  \mathbf{X} \equiv
  \begin{bmatrix}
    \bm{x}_{t_0}^T & 1 \\
    \vdots     & \vdots \\
    \bm{x}_{t_N}^T & 1 \\
  \end{bmatrix}
\end{equation}
where $\mathbf{D}$ ($\in \mathbb{C}^{N \times N_\mathrm{out}}$) is the matrix of teacher signals corresponding to $\mathbf{X}$, and $\mathbf{X}^*$ is Hermitian conjugate of $\mathbf{X}$.

CVRC has high generalization ability in the complex domain, that is, the ability to deal with data obtained under different conditions as well as real observation data including noise. The proposed method is capable of processing interferograms including much noise and/or obtained by other observations.

\section{Experiment on aspect classification}
\label{s:aspect}

\subsection{Data used in the experiment}
\label{ss:data}

\begin{figure}[tp]
  \centering
  \begin{minipage}{0.45\linewidth}
    \includegraphics[width=0.98\linewidth]{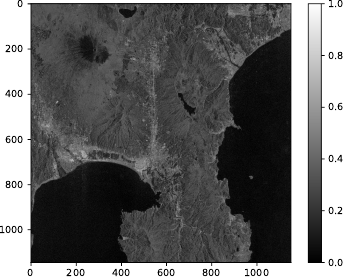}
    \subcaption{}
  \end{minipage}
  \begin{minipage}{0.45\linewidth}
    \includegraphics[width=0.98\linewidth]{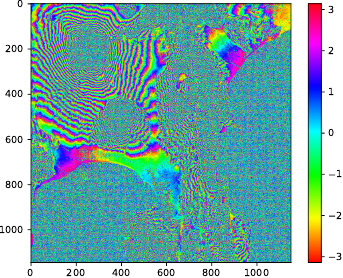}
    \subcaption{}
  \end{minipage}

  \caption{Original complex interferogram data : (a) amplitude and (b) phase images around Mt. Fuji}
  \label{fig:img-origin}
\end{figure}

\begin{figure}[tp]
  \centering
  \begin{minipage}{0.45\linewidth}
    \includegraphics[width=0.98\linewidth]{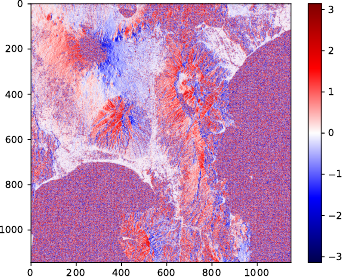}
    \subcaption{}
  \end{minipage}
  \begin{minipage}{0.45\linewidth}
    \includegraphics[width=0.98\linewidth]{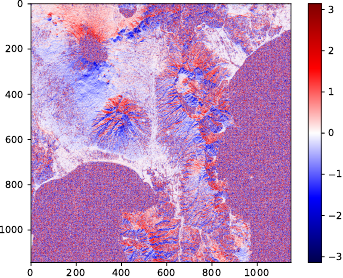}
    \subcaption{}
  \end{minipage}

  \caption{Spatial differences of phase data : (a) east-west and (b) north-south directions around Mt. Fuji}
  \label{fig:img-diff}
\end{figure}

We deal with an interferogram obtained from two JAXA ALOS data around Mt. Fuji observed on November 25, 2010 and April 12, 2011.
We conducted preprocessing of removing orbital fringes, $16\times 8$ multilooking and normalizing log amplitude in such a manner that the maximum equals to unity and that the noise-equivalent backscattering coefficient is zero.
The resolution after this preprocessing becomes 30~m/pixel (range) $\times$ 50~m/pixel (azimuth).
Fig.~\ref{fig:img-origin} shows the interferogram in the normalized amplitude and phase.
We generate the phase difference data of east-west (range) and north-south (azimuth) directions as shown in Fig.~\ref{fig:img-diff}, in order to handle phase changes in space directly without unwrapping to avoid unwrapping errors.
The amplitude data remains without changes.

\subsection{Learning and estimation procedure}

\begin{figure*}[tp]
  \centering
  \includegraphics[width=0.98\linewidth]{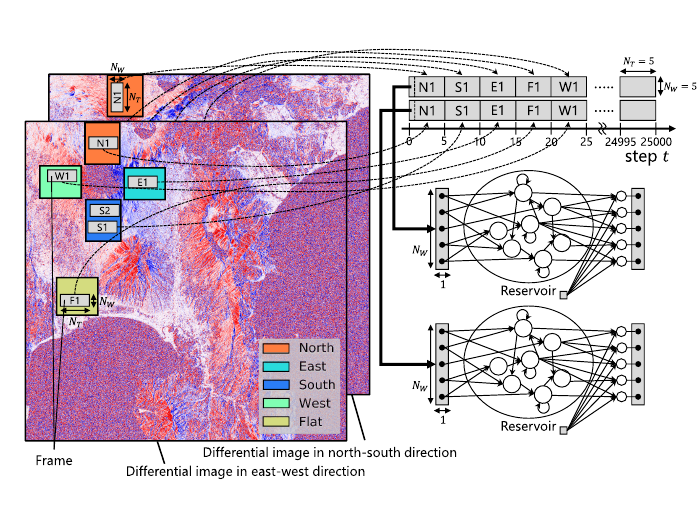}
  \caption{Serialization of the two-dimensional data in the learning process.}
  \label{fig:learning}
\end{figure*}

Fig.~\ref{fig:learning} presents the total learning process.
First, we delimit teacher signal areas from the two phase difference images and assign labels of 5 classes: north, east, south and west slopes and flat plane.
Next, we cut out small teacher frames randomly in these areas and arrange them to make sequential data. 
The sizes of the frames are $N_W\times N_T$ and $N_T\times N_W$ pixels for east-west and north-south difference images, respectively.
Then, we make long sequential data by aligning 1000 frames where $N_W = 5$ and $N_T = 5$ in each area.
Finally, we feed them to the reservoirs in turn.

We decide the number of neurons in the reservoirs $N_\mathrm{res} = 5$ and the number of neurons in the output layers $N_\mathrm{out} = 5$ corresponding to the number of the labels for the classification.
We set the spectral radius of the initial recurrent weight matrices $\sigma(\mathbf{W}_\mathrm{res})$, the desired radius $\sigma_\mathrm{d}$ and the leaking rate $\alpha$ at $0.16$, $0.10$ and $0.30$, respectively such that the magnitude of data once fed to the reservoir decreases to $1/N_T$ in $N_T$ time steps.
The teacher matrix $\mathbf{D}$ is constructed corresponding to the input teacher-frame sequence as
\begin{equation}
  \label{formula:desire}
  \mathbf{D}=
  \overbrace{ 
  \left[
  \begin{array}{rrrrr}
     1  & -1  & -1 & \cdots & -1 \\
    -1  & -1  &  1 & \cdots & -1 \\
    -1  &  1  & -1 & \cdots & -1 \\
    -1  & -1  & -1 & \cdots &  1 \\
    -1  & -1  & -1 & \cdots & -1
  \end{array}
  \right]
  }^{N}
  \begin{array}{l}
    \leftarrow \mathrm{North} \\
    \leftarrow \mathrm{East} \\
    \leftarrow \mathrm{South} \\
    \leftarrow \mathrm{West} \\
    \leftarrow \mathrm{Flat}
  \end{array}
\end{equation}
We calculate $\mathbf{W}_\mathrm{out}$ by (\ref{formula:learning}) with a regularization parameter $\lambda$ of $10^{-12}$.

\begin{figure*}[tp]
  \centering
  \includegraphics[width=0.98\linewidth]{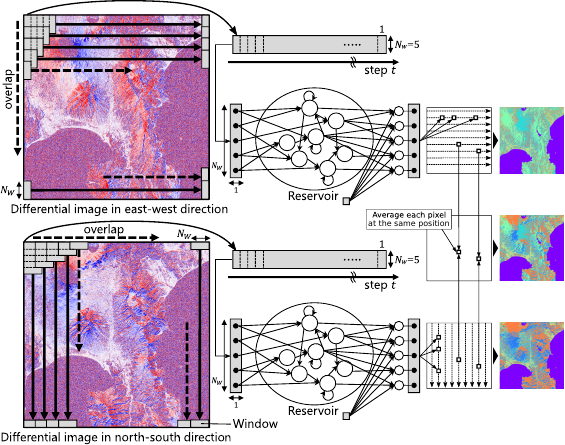}
  \caption{Serialization of the two-dimensional data in the classification process.}
  \label{fig:estimation}
\end{figure*}

Fig.~\ref{fig:estimation} presents the total classification process.
We first prepare east-west and north-south difference images for estimation.
Secondly, we generate two sequential data by using these images as follows.
We scan the east-west difference image from upper left rightward with a window of $N_W\times 1$ pixels.
Having scanned the image to the right end, then we shift the window down by a single pixel and scan the image to the right end again.
We scan the north-south difference image from upper left downward with a window of $1\times N_W$ pixels.
After scanning the image to the lower end, we shift the window rightward by a single pixel and scan to the lower end.
We repeat these processes to the lower right to make sequential data.

Then, we input each sequential data made from east-west and north-south difference images to the networks learned with each image and obtain sequential outputs.
Finally, we average these two outputs.
The area corresponding to the neuron closest to unity is the classification result for the input.

TABLE~\ref{tab:params} lists hyperparameters in the experiment.
We decided $N_W$ and $N_T$ by the reasons described later.
We also use NumPy of python library in the experiment.

\begin{table}[tp]
  \caption{Hyperparameters for the serialization process and the CVRC network in the experiment on aspect classification}
  \label{tab:params}
  \centering
  \begin{tabular}{rlr}
    \hline
    \multicolumn{2}{c}{Parameter} & \multicolumn{1}{c}{Value} \\
    \hline \hline
    Width of frame                            & $N_W$               & 5 \\
    Height of frame                           & $N_T$               & 5 \\
    Sequence length                           & $N$                 & 5,000\\
    The number of neurons in the input layer  & $N_\mathrm{in}$     & 5 \\
    The number of neurons in the reservoir    & $N_\mathrm{res}$    & 5 \\
    The number of neurons in the output layer & $N_\mathrm{out}$    & 5 \\
    Desireble spectral radius                 & $\sigma_\mathrm{d}$ & 0.10 \\
    Leaking rate                              & $\alpha$            & 0.30 \\
    Regularization parameter                  & $\lambda$           & $10^{-12}$ \\
    \hline
  \end{tabular}
\end{table}

\subsection{The ground truth and neighbor difference method}
\label{ss:truth_simple}

To evaluate the experimental results, we generate a ground-truth phase data by using a digital elevation data provided by Geospatial Information Authority of Japan (GSI).
The phase data and the interferogram used in the experiment have the identical resolution. 
After calculating the difference values of two adjacent pixels in east-west and north-south directions, we classify the difference data to the four cardinal directions.
The threshold between flat and slope is set at the intermediate value between mean values in four slope areas and flat area used as the teacher areas.
We also mask out pixels corresponding to water areas obtained from the GSI data to exclude the sea and lake areas in the evaluation.

Additionally, we calculate the spatial difference of neighboring pixels in the interferogram phase map.
Hereinafter, we call the method neighbor difference method.

\subsection{Experimental results}
\label{ss:result}

\subsubsection{Classification results for east-west data, north-south data and their integration}

\begin{figure*}[tp]
  \centering
  \begin{minipage}{0.45\linewidth}
    \includegraphics[width=0.98\linewidth]{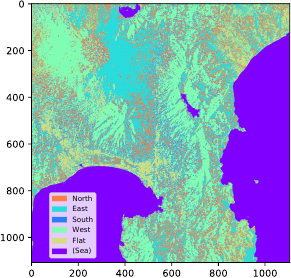}
    \subcaption{}
  \end{minipage}
  \begin{minipage}{0.45\linewidth}
    \includegraphics[width=0.98\linewidth]{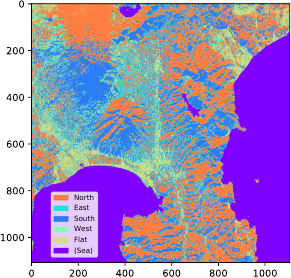}
    \subcaption{}
  \end{minipage}\\

  \begin{minipage}{0.45\linewidth}
    \includegraphics[width=0.98\linewidth]{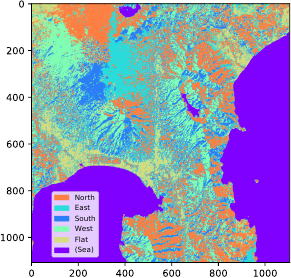}
    \subcaption{}
  \end{minipage}
  \begin{minipage}{0.45\linewidth}
    \includegraphics[width=0.98\linewidth]{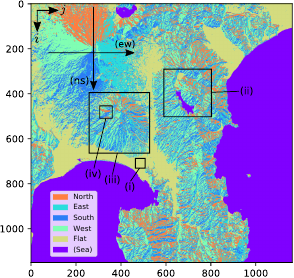}
    \subcaption{}
  \end{minipage}
  \caption{CVRC classification results for (a) east-west difference data, (b) north-south difference data, (c) integration of these two outputs, and (d) the ground truth.}
  \label{fig:est}
\end{figure*}

Fig.~\ref{fig:est} presents CVRC classification results for (a) the east-west difference data, (b) the north-south difference data, (c) the integration of these two outputs and (d) the ground truth.
The result by using east-west difference data emphasizes east and west slopes clearly.
In other words, north and south slopes are regarded as east or west in error.
This originates from the anisotropy in the differentiation and scanning.
Similarly, the result by using north-south difference data enhances north and south slopes.
In contrast, the integration result in Fig.~\ref{fig:est}{(c)} presents a good aspect decision similar to the ground truth shown in Fig.~\ref{fig:est}{(d)}.

\subsubsection{Comparison with conventional methods}

We conducted the classification by using CVRC, RVRC, CVCNN\cite{Sunaga2019}, and neighbor difference method to compare their estimation performance and computational cost.
We show classification results and its accuracy of the whole area, flat plane, periphery of Lake Ashi, Mt. Ashitaka, and west ridge of Mt. Ashitaka in order to elucidate classification features in the respective methods.

\begin{figure*}[tp]
  \centering
  \begin{minipage}{0.19\linewidth}
    \includegraphics[width=0.98\linewidth]{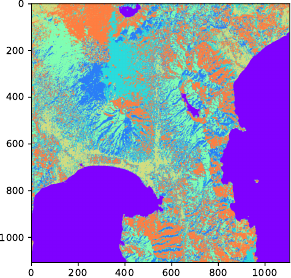}
    \subcaption*{(a-1)}
  \end{minipage}
  \begin{minipage}{0.19\linewidth}
    \includegraphics[width=0.98\linewidth]{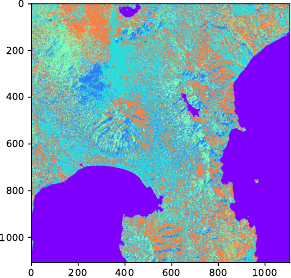}
    \subcaption*{(a-2)}
  \end{minipage}
  \begin{minipage}{0.19\linewidth}
    \includegraphics[width=0.98\linewidth]{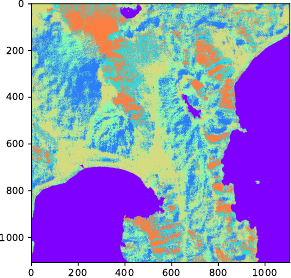}
    \subcaption*{(a-3)}
  \end{minipage}
  \begin{minipage}{0.19\linewidth}
    \includegraphics[width=0.98\linewidth]{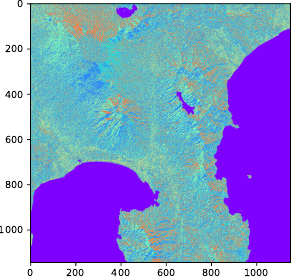}
    \subcaption*{(a-4)}
  \end{minipage}
  \begin{minipage}{0.19\linewidth}
    \includegraphics[width=0.98\linewidth]{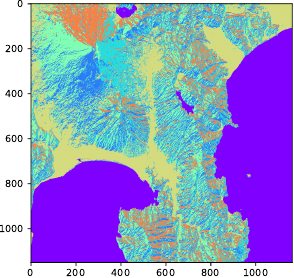}
    \subcaption*{(a-5)}
  \end{minipage}\\

  \begin{minipage}{0.19\linewidth}
    \includegraphics[width=0.98\linewidth]{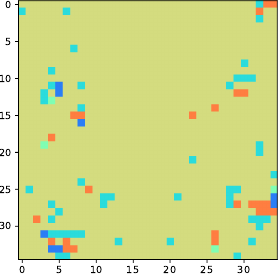}
    \subcaption*{(b-1)}
  \end{minipage}
  \begin{minipage}{0.19\linewidth}
    \includegraphics[width=0.98\linewidth]{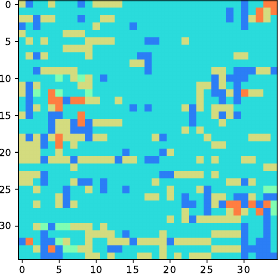}
    \subcaption*{(b-2)}
  \end{minipage}
  \begin{minipage}{0.19\linewidth}
    \includegraphics[width=0.98\linewidth]{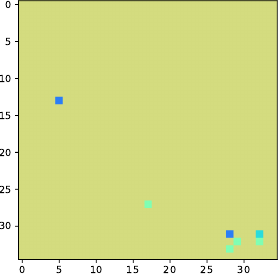}
    \subcaption*{(b-3)}
  \end{minipage}
  \begin{minipage}{0.19\linewidth}
    \includegraphics[width=0.98\linewidth]{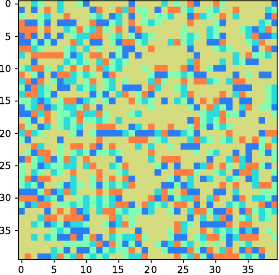}
    \subcaption*{(b-4)}
  \end{minipage}
  \begin{minipage}{0.19\linewidth}
    \includegraphics[width=0.98\linewidth]{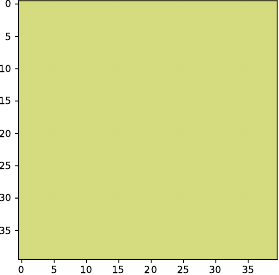}
    \subcaption*{(b-5)}
  \end{minipage}\\

  \begin{minipage}{0.19\linewidth}
    \includegraphics[width=0.98\linewidth]{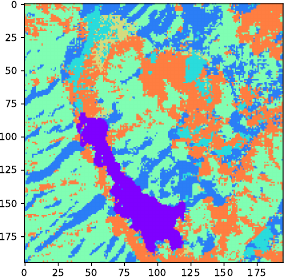}
    \subcaption*{(c-1)}
  \end{minipage}
  \begin{minipage}{0.19\linewidth}
    \includegraphics[width=0.98\linewidth]{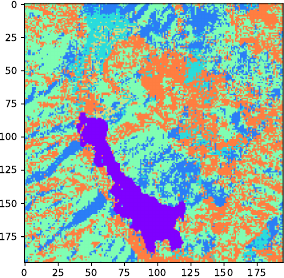}
    \subcaption*{(c-2)}
  \end{minipage}
  \begin{minipage}{0.19\linewidth}
    \includegraphics[width=0.98\linewidth]{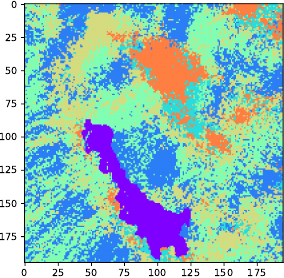}
    \subcaption*{(c-3)}
  \end{minipage}
  \begin{minipage}{0.19\linewidth}
    \includegraphics[width=0.98\linewidth]{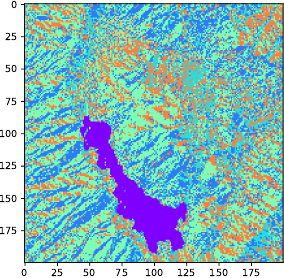}
    \subcaption*{(c-4)}
  \end{minipage}
  \begin{minipage}{0.19\linewidth}
    \includegraphics[width=0.98\linewidth]{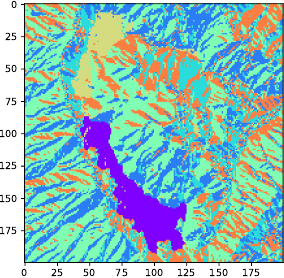}
    \subcaption*{(c-5)}
  \end{minipage}\\

  \begin{minipage}{0.19\linewidth}
    \includegraphics[width=0.98\linewidth]{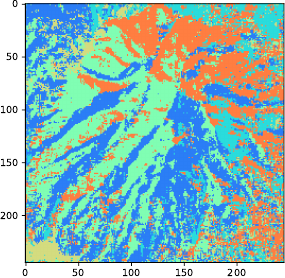}
    \subcaption*{(d-1)}
  \end{minipage}
  \begin{minipage}{0.19\linewidth}
    \includegraphics[width=0.98\linewidth]{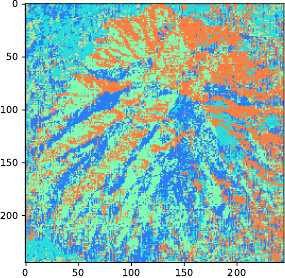}
    \subcaption*{(d-2)}
  \end{minipage}
  \begin{minipage}{0.19\linewidth}
    \includegraphics[width=0.98\linewidth]{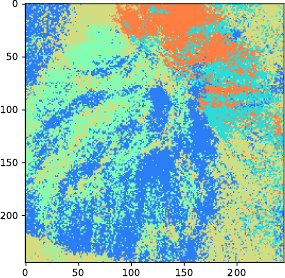}
    \subcaption*{(d-3)}
  \end{minipage}
  \begin{minipage}{0.19\linewidth}
    \includegraphics[width=0.98\linewidth]{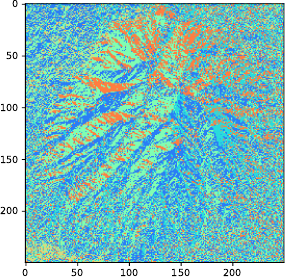}
    \subcaption*{(d-4)}
  \end{minipage}
  \begin{minipage}{0.19\linewidth}
    \includegraphics[width=0.98\linewidth]{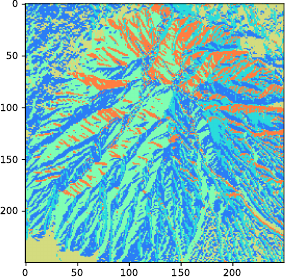}
    \subcaption*{(d-5)}
  \end{minipage}\\

  \begin{minipage}{0.19\linewidth}
    \includegraphics[width=0.98\linewidth]{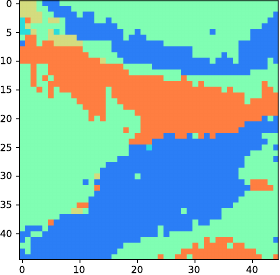}
    \subcaption*{(e-1)}
  \end{minipage}
  \begin{minipage}{0.19\linewidth}
    \includegraphics[width=0.98\linewidth]{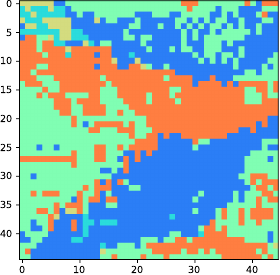}
    \subcaption*{(e-2)}
  \end{minipage}
  \begin{minipage}{0.19\linewidth}
    \includegraphics[width=0.98\linewidth]{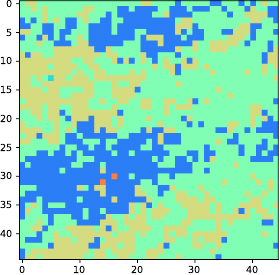}
    \subcaption*{(e-3)}
  \end{minipage}
  \begin{minipage}{0.19\linewidth}
    \includegraphics[width=0.98\linewidth]{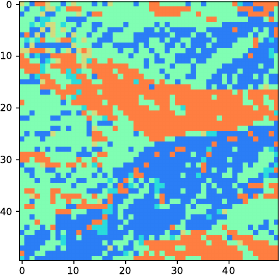}
    \subcaption*{(e-4)}
  \end{minipage}
  \begin{minipage}{0.19\linewidth}
    \includegraphics[width=0.98\linewidth]{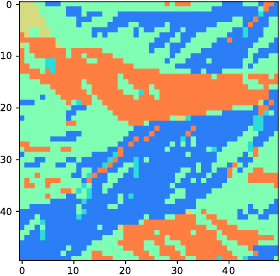}
    \subcaption*{(e-5)}
  \end{minipage}

  \caption{Classification results in (a-$\star$) whole area, (b-$\star$) flat plane, (c-$\star$) periphery of Lake Ashi, (d-$\star$) Mt. Ashitaka, and (e-$\star$) a west ridge of Mt. Ashitaka by using ($\star$-1) CVRC, ($\star$-2) RVRC, ($\star$-3) CVCNN, and ($\star$-4) neighbor difference method as well as ($\star$-5) ground truth.}
  \label{fig:est_detail}
\end{figure*}

Fig.~\ref{fig:est_detail} shows classification results for the 5 areas: 
(a-$\star$) whole area, 
(b-$\star$) a part of flat plane in area~(i) ($i=690\textup{--}730, j=460\textup{--}500$), 
(c-$\star$) periphery of Lake Ashi in area~(ii) ($i=300\textup{--}500, j=600\textup{--}800$), 
(d-$\star$) Mt. Ashitaka in area~(iii) ($i=400\textup{--}670, j=260\textup{--}530$), and 
(e-$\star$) a west ridge of Mt. Ashitaka in area~(iv) ($i=460\textup{--}510, j=310\textup{--}360$)
obtained by using ($\star$-1) CVRC, ($\star$-2) RVRC, ($\star$-3) CVCNN, ($\star$-4) neighbor difference method, and ($\star$-5) the ground truth.
TABLE~\ref{tab:compare} shows accuracies, learning time, and classification time.
We define the accuracy as the concordance rate between classification results and the ground truth.

The flat plane is easily classifiable because of its almost constant amplitude and phase values and the resulting low spatial frequency.
RVRC, having low generalization ability in the complex domain, shows a low accuracy of 17.8\%.
However, CVCNN and CVRC, which deal with complex values consistently, present high accuracies of 99.4\% and 93.1\%, respectively, as shown in TABLE~\ref{tab:compare} {(i)}.

In contrast, periphery of Lake Ashi and Mt. Ashitaka contain complex shapes with high spatial frequencies, which leads to difficulty in classification.
Then, basically all the methods present low accuracies of 51.9\%, 56.6\%, 57.0\%, and 64.3\% in TABLE~\ref{tab:compare} {(ii)--(iv)}.

Next, we compare the classification performances of CVRC and CVCNN.
CVRC is inferior to CVCNN for the flat plane.
But, in Figs. \ref{fig:est_detail}{($\star$-1)}, {($\star$-3)} and {($\star$-4)}, it can classify small ridges of the periphery of Lake Ashi and Mt. Ashitaka well compared with CVCNN.
TABLE~\ref{tab:compare} presents that CVRC achieved a higher accuracy than CVCNN in the whole area.
In terms of the computational cost, CVRC requires only about one-hundredth learning time and one-fifth classification processing time compared with CVCNN.

We also compare CVRC and RVRC.
Figs.~\ref{fig:est_detail}{($\star$-2)}, {($\star$-3)} and {($\star$-4)} shows that both CVRC and RVRC classify small ridges in the periphery of Lake Ashi and Mt. Ashitaka very well.
In the flat plane, CVRC presents a better result containing less salt-and-pepper noise totally than RVRC.
TABLE~\ref{tab:compare} reveals that CVRC achieves a higher accuracy than RVRC for the whole area.
CVRC and RVRC consume the same calculation time.

\begin{table*}[tp]
  \caption{Comparison of accuracy and calculation time between CVRC, RVRC, CVCNN, and neighbor difference method}
  \centering
  \scalebox{0.63}{
  \begin{tabular}{c|ccccc|cc}
    \hline
    \multirow{3}{*}{Method} & \multicolumn{5}{|c|}{Accuracy~[\%]} & \multicolumn{2}{|c}{Calculation Time~[sec]} \\ \cline{2-8}
    & \multirow{2}{*}{(i)~Flat} & (ii)~Periphery of & \multirow{2}{*}{(iii)~Mt. Ashitaka} & (iv)~Mt. Ashitaka & \multirow{2}{*}{Overall} & Learning & Classification \\ 
    & & ~~~~~Lake Ashi & & ~~~~~(west-ridge) & & Time & Time \\ 
    \hline \hline
    CVRC    & 93.1          & \textbf{56.2} & \textbf{48.6} & \textbf{65.6} & \textbf{64.3} & \textbf{6}   & \textbf{300}   \\
    RVRC    & 17.8          & 50.4          & 40.7          & 60.8          & 57.0          & \textbf{6}   & \textbf{300}   \\
    CVCNN   & \textbf{99.4} & 33.9          & 31.1          & 26.8          & 56.6          & 660 & 1440  \\
    Simple difference  & \multirow{1}{*}{42.4} & \multirow{1}{*}{38.8} & \multirow{1}{*}{32.5} & \multirow{1}{*}{43.6} & \multirow{1}{*}{51.9} & \multirow{1}{*}{---} & \multirow{1}{*}{---} \\
    \hline
  \end{tabular}
  }
  \label{tab:compare}
\end{table*}

\subsubsection{Application of learning to another region in another data}
\begin{figure}[tp]
  \centering
  \begin{minipage}{0.45\linewidth}
    \includegraphics[width=0.98\linewidth]{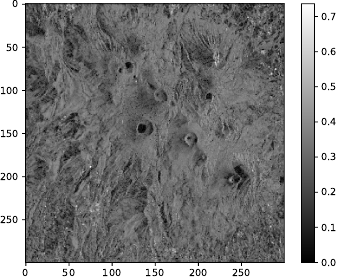}
    \subcaption{}
  \end{minipage}
  \begin{minipage}{0.45\linewidth}
    \includegraphics[width=0.98\linewidth]{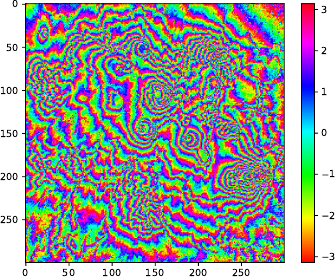}
    \subcaption{}
  \end{minipage}\\%
  \begin{minipage}{0.45\linewidth}
    \includegraphics[width=0.98\linewidth]{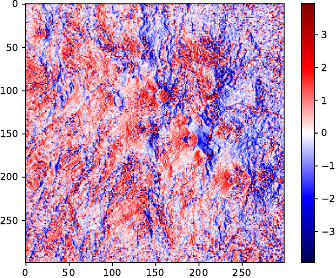}
    \subcaption{}
  \end{minipage}
  \begin{minipage}{0.45\linewidth}
    \includegraphics[width=0.98\linewidth]{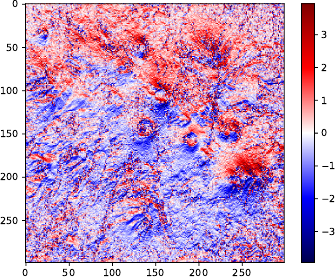}
    \subcaption{}
  \end{minipage}
  \caption{Original complex interferogram data : (a) amplitude and (b) phase images and spatial differences of phase data : (a) east-west and (b) north-south directions around Shinmoe-dake.}
  \label{fig:img-origin-shinmoe}
\end{figure}

To validate generalization characteristics of the CVRC, we conduct an experiment using the observation data that is different from the data used for learning.
The interferogram used in this experiment is obtained from two JAXA ALOS data around Shinmoe-dake observed on April 14, 2009 and May 30, 2009.
Figs.~\ref{fig:img-origin-shinmoe}{(a)} and {(b)} show the interferogram in normalized log-amplitude and phase.
We generated the difference phase data of east-west (range) and north-south (azimuth) directions, as shown in Figs.~\ref{fig:img-origin-shinmoe}{(c)} and {(d)}.
We also equalized the height ambiguity of this interferogram to the interferogram around Mt.Fuji in \ref{ss:data} as preprocessing.
The resolution and estimation method are the same as those in the previous section.

\begin{figure}[tp]
  \centering
  \begin{minipage}{0.45\linewidth}
    \includegraphics[width=0.98\linewidth]{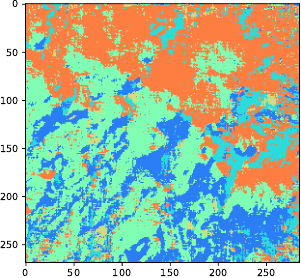}
    \subcaption{}
  \end{minipage}
  \begin{minipage}{0.45\linewidth}
    \includegraphics[width=0.98\linewidth]{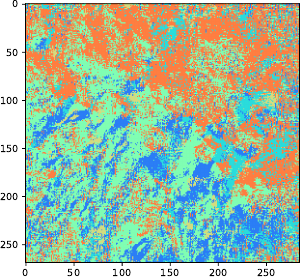}
    \subcaption{}
  \end{minipage}\\%
  \begin{minipage}{0.45\linewidth}
    \includegraphics[width=0.98\linewidth]{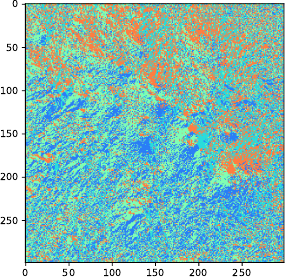}
    \subcaption{}
  \end{minipage}
  \begin{minipage}{0.45\linewidth}
    \includegraphics[width=0.98\linewidth]{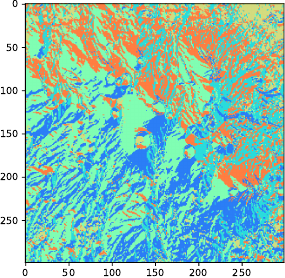}
    \subcaption{}
  \end{minipage}
  \caption{Classification results in Shinmoe-dake by using (a) CVRC, (b) RVRC, (c), and neighbor difference method as well as (d) ground truth.}
  \label{fig:est-shinmoe}
\end{figure}

Fig.~\ref{fig:est-shinmoe} shows classification results obtained by using {(a)} CVRC, {(b)} RVRC, {(c)} neighbor difference method, and {(d)} the ground truth.
The classification accuracies are 37.1\%, 33.6\% and 32.1\% by using CVRC, RVRC and neighbor difference method, respectively.
Despite the fact that this region containing many small mountains is difficult for classification, CVRC is better than other methods in spite of the difference in region and data.
Then we found that CVRC has a high generalization ability.

\subsection{Discussion}
\label{ss:discussion}

\subsubsection{Comparison with conventional methods}

\begin{figure}[tp]
  \centering
  \begin{minipage}{0.8\linewidth}
    \includegraphics[width=0.98\linewidth]{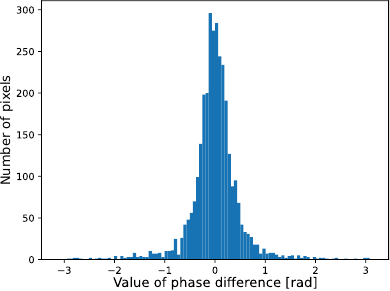}
    \subcaption{}
  \end{minipage}\\
  \begin{minipage}{0.8\linewidth}
    \includegraphics[width=0.98\linewidth]{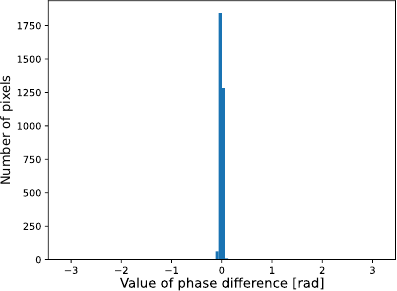}
    \subcaption{}
  \end{minipage}

  \caption{Histgrams of the phase difference data obtained from (a) the interferogram and (b) the ground truth in the flat area.}
  \label{fig:hist}
\end{figure}

First, we discuss the classification result of neighbor difference method.
Fig.~\ref{fig:hist} presents histgrams of the phase difference data of {(a)} the interferogram and {(b)} the ground truth in the flat area.
These variances $\sigma^2$ are $2.47\times10^{-1}$ and $5.59\times10^{-4}$.
Given this histgrams, the interferogram contains much more noise.
Simple difference method that just differences two adjacent pixels shows totally a low accuracy due to the low noise robustness.
In contrast, CVRC is able to reduce noise effect because the short-term memory in the RC works to mitigate fluctuation in the input signals.

Next, we discuss the classification performance of CVCNN.
In general, CVCNN is suitable for learning area shapes in its convolutional process to extract shape features and in the pooling process to weaken the influence of translation, rotation and scaling.
However, the slope orientation is independent of area shapes.
Though CVCNN can classify targets like flat plane and slopes having only low spatial frequency, it has difficulty in classifying targets with high spatial frequency such as slopes containing small ridges.
This is because CVCNN captures window features.
In contrast, CVRC catches small changes by dealing with the two-dimensional data as sequential pixel data to generate classification results with high resolution.

Next, we discuss the classification performance of RVRC.
RVRC sometimes fails in classification of data represented as amplitude and phase information.
Generally, flat planes feature large amplitude and constant phase.
Slopes also present phase changes meaningfully.
Then, RVRC shows a lower performance with more salt-and-pepper noise totally.

In addition, we discuss neuron signals in the reservoir and output error (root mean square error : RMSE) in the estimation to elucidate the classification process in RVRC and CVRC.
We define RMSE as
\begin{equation}
  \label{formula:error}
  \mathrm{RMSE} = \sqrt{\frac{1}{K}\sum^K_{k=1} |y_k-d_k|^2}
\end{equation}
where $y_k$ is the signal of an output neuron, $d_k$ is the desirable output signal, and $K$ is the number of classes.
The RMSE is mostly over 1 for an erroneous decision.

\begin{figure}[tp]
  \centering
  \includegraphics[width=0.98\linewidth]{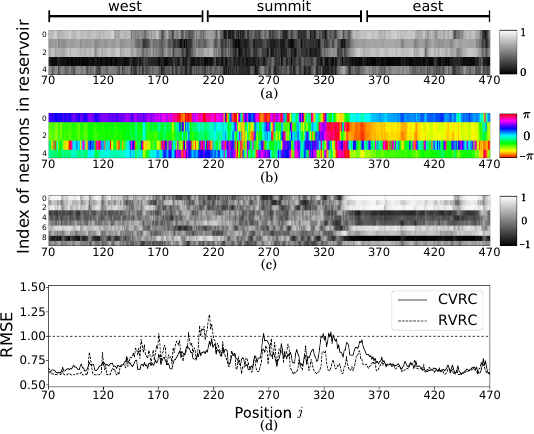}
  \caption{(a) Amplitude and (b) phase of the signals in the reservoir of CVRC, (c) the signals in the reservoir of RVRC, and (d) RMSE of CVRC and RVRC in scanning of the area {(ew)} ($i=210, j=70\textup{--}471$) from the west slope (Position $j=70\textup{--}210$) passing the summit to the east slope (Position $j=360\textup{--}470$) of Mt. Fuji.}
  \label{fig:res_ew}
\end{figure}

Fig.~\ref{fig:res_ew} shows (a) amplitude and (b) phase of the signals in the reservoir of CVRC, (c) the signals in the reservoir of RVRC, and (d) RMSE of CVRC and RVRC in the scanning of the area {(ew)} ($i=210, j=70\textup{--}471$) from the west slope (Position $j=70\textup{--}210$) passing the summit to the east slope (Position $j=360\textup{--}470$) of Mt. Fuji.
The periphery of $j=280$ is the summit of Mt. Fuji.
According to the diagrams, CVRC has almost the same amplitude and difference constant phase values in east and west slope areas, respectively.
Therefore, CVRC classifies east and west slopes based on the phase in the reservoir neurons.
On the other hand, RVRC has similar neuron values in the west and east areas.
Hence, the classification in the RVRC is difficult.
Though both RMSE are similar to each other in the east slope, RVRC has a larger RMSE in the west slope than CVRC.

Around the summit of Mt. Fuji, the amplitude is small and the phase is almost random because of the scree caused multiple scattering.
Hence, the reservoir signals of the CVRC are very small in the amplitude and at random in the phase, and RVRC has random reservoir signals as well.

\begin{figure}[tp]
  \centering
  \includegraphics[width=0.98\linewidth]{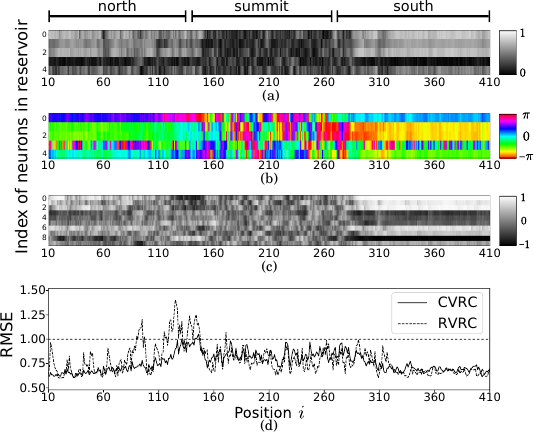}
  \caption{(a) Amplitude and (b) phase of the signals in the reservoir of CVRC, (c) the signals in the reservoir of RVRC, and (d) RMSE of CVRC and RVRC in scanning of the area~{(ns)} ($i=10\textup{--}411, j=270$) from the north slope (Position $i=10\textup{--}150$) passing the summit to the south slope (Position $i=270\textup{--}410$) of Mt. Fuji.}
  \label{fig:res_ns}
\end{figure}

Fig.~\ref{fig:res_ns} shows (a) amplitude and (b) phase of the signals in the reservoir of CVRC, (c) the signals in the reservoir of RVRC, and (d) RMSE of CVRC and RVRC in scanning of the area~{(ns)} ($i=10\textup{--}411, j=270$) from the north slope (Position $i=10\textup{--}150$) to the south slope (Position $i=270\textup{--}410$) of Mt. Fuji.
At around $i=210$ is the summit of Mt. Fuji.
The data for north and south slopes indicates almost the same as those for east and west slopes.
Then, although CVRC classifies north and south slopes from phase information successfully, classification in RVRC fails because the signals of north and south slopes have similar values and the north slope signals fluctuate at random.

The phase of CVRC represents a feature more clearly than that of RVRC.
In Fig.~\ref{fig:res_ns}{(d)}, the output error of CVRC is smaller than that of the RVRC, showing good utilization of the input signals outside the noisy summit.

\subsubsection{Performance dependency on the reservoir neuron number}

\begin{figure}[tp]
  \centering
  \includegraphics[width=0.98\linewidth]{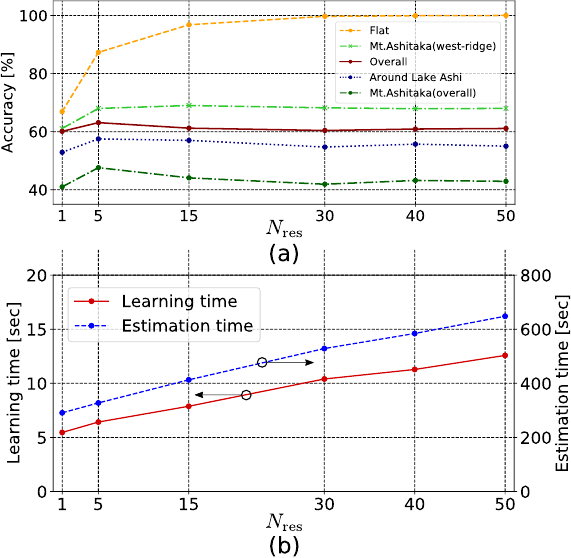}
  \caption{(a) Accuracy and (b) learning and classification time versus the number of neurons in the reservoir $N_\mathrm{res}$.}
  \label{fig:res_nodes}
\end{figure}

We discuss the influence of the number of neurons in the reservoir on the classification results.
Fig~\ref{fig:res_nodes} shows (a) accuracy for Areas {(i)--(iv)} and for whole area, and (b) required learning and classification time when the number of neurons in the reservoir $N_\mathrm{res}$ is chosen as $1$, $5$, $15$, $30$, $40$, and $50$.
We find in Fig.~\ref{fig:res_nodes}{(a)} that the more the number of neurons is, the higher the accuracy becomes. The accuracy is almost $100\%$ when $N_\mathrm{res}\geq 30$ in the flat plane.
Periphery of Lake Ashi, Mt. Ashitaka, and whole area shows the highest accuracies when $N_\mathrm{res}$ is 5.
That is, a higher $N_\mathrm{res}$ does not always results in a high accuracy. 
According to Fig.~\ref{fig:res_nodes}{(b)}, learning and classification time increase linearly as the neuron number grows.
By considering this trade-off, we decided $N_\mathrm{res}=5$.

\subsubsection{Performance dependency on the frame size}

\begin{figure*}[tp]
  \centering
  \includegraphics[width=0.98\linewidth]{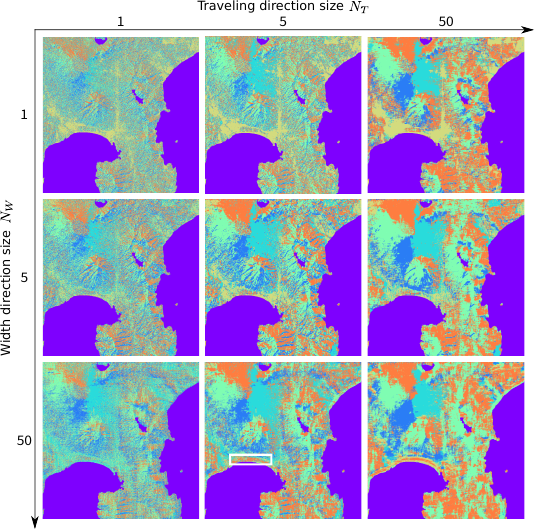}
  \caption{Comparison of classification results for various frame sizes.}
  \label{fig:frame}
\end{figure*}

We discuss the influence of the frame size on the classification results.
Fig.~\ref{fig:frame} shows CVRC results for the size in width direction $N_W=1$, $5$, or $50$ and the size in traveling direction $N_T=1$, $5$, or $50$.
When we compare these results with the ground truth (Fig.~\ref{fig:est}{(d)}), we find the condition that $N_W=5$ and $N_T=5$ presents the best performance.

The classification result for $N_W=1$ and $N_T=1$ contains high salt-and-pepper noise compared to the ground truth in Fig.~\ref{fig:est}{(d)}.
We also find that the result for $N_W=50$ and $N_T=50$ shows degraded resolution and contains horizontal and vertical artifact lines.

Next, we examine the influence of $N_W$ and $N_T$.
When $N_W$ is large, the linear combination (averaging) of the input signals $\mathbf{W}_\mathrm{in}\bm{u}_t$ loses high frequency local information.
The results for $N_W=50$ and $N_T=5$ shows erroneous north slope for a flat area (white rectangle) because of the averaging effect similar to the above one, resulting in smaller amplitude inconsistent with typical flat area having intense scattering.

The classification result for $N_W=1$ and $N_T=50$ shows a degraded resolution but presents a good result similar to that for $N_W=5$ and $N_T=5$.
This is because the current input signals have the most influence and the past input signals vanish exponentially by the dynamics of the reservoir and construct the output appropriately after the learning.
Thereby, the larger the size of traveling direction is, the less salt-and-pepper noise becomes.

\section{Experiment on slope angle estimation}
\label{s:slope}

The experiment of aspect classification in the previous section is a discrete task.
In this section, we conduct an experiment on east-west slope angle estimation, which is more difficult.

\subsection{Experimental setup}

The interferogram used here is identical with that in \ref{ss:data}.
The ground truth and the neighbor-difference estimation are generated by the procedure explained in \ref{ss:truth_simple}.

\begin{figure*}[tp]
  \centering
  \includegraphics[width=0.98\linewidth]{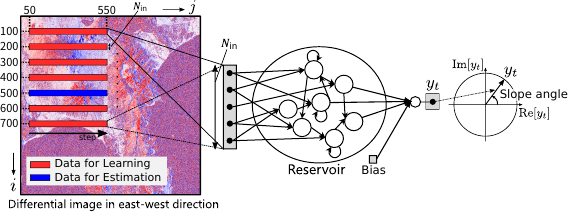}
  \caption{Procedure of the experiment on slope angle estimation for learning and estimation.}
  \label{fig:method-slope}
\end{figure*}

Fig.~\ref{fig:method-slope} shows the procedure of learning and estimation in the slope angle estimation task.
The input data fed to the CVRC network is one of the east-west difference data.
We train the network by feeding input data at $i=100,200,300,400,600,700$ and $j=50\textup{--}550$ with a $d$-pixel past teacher signals to the output.
We also estimate the network using the data used for learning at $i=300$ and $j=50\textup{--}550$ as well as a new data running through the center of Mt. Ashitaka at $i=500$ and $j=50\textup{--}550$.
Table.~\ref{tab:params-slope} lists hyperparameters in the experiment.
The number of neurons in the reservoir is larger than those in the aspect classification because we estimate continuous values quantitatively in this experiment.

\begin{table}[tp]
  \caption{Hyperparameters for the CVRC network in the experiment on slope angle estimation}
  \label{tab:params-slope}
  \centering
  \begin{tabular}{rlr}
    \hline
    \multicolumn{2}{c}{Parameter} & \multicolumn{1}{c}{Value} \\
    \hline \hline
    The number of neurons in the input layer  & $N_\mathrm{in}$     & 5 \\
    The number of neurons in the reservoir    & $N_\mathrm{res}$    & 300 \\
    The number of neurons in the output layer & $N_\mathrm{out}$    & 1 \\
    Desireble spectral radius                 & $\sigma_\mathrm{d}$ & 0.90 \\
    Leaking rate                              & $\alpha$            & 0.80 \\
    Regularization parameter                  & $\lambda$           & $10^{-12}$ \\
    Delay                                     & $d$                 & 5 \\
    \hline
  \end{tabular}
\end{table}

\subsection{Results}

\begin{figure}[tp]
  \centering
  \begin{minipage}{0.45\linewidth}
    \includegraphics[width=0.98\linewidth]{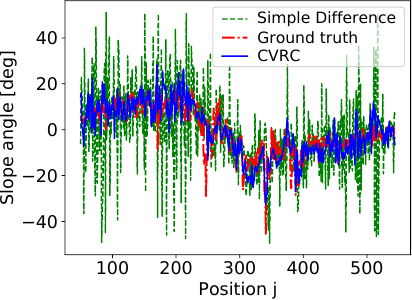}
    \subcaption*{(a-1)}
  \end{minipage}
  \begin{minipage}{0.45\linewidth}
    \includegraphics[width=0.98\linewidth]{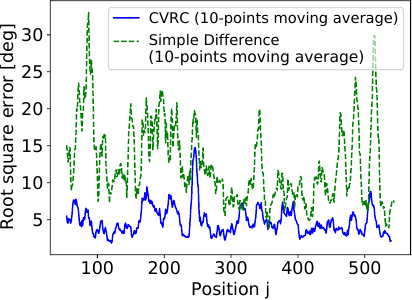}
    \subcaption*{(a-2)}
  \end{minipage} \\

  \begin{minipage}{0.45\linewidth}
    \includegraphics[width=0.98\linewidth]{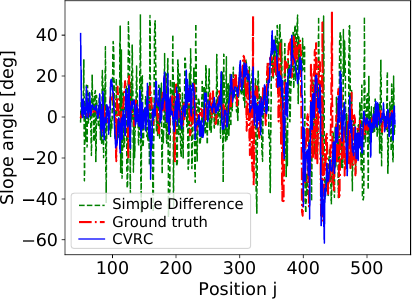}
    \subcaption*{(b-1)}
  \end{minipage}
  \begin{minipage}{0.45\linewidth}
    \includegraphics[width=0.98\linewidth]{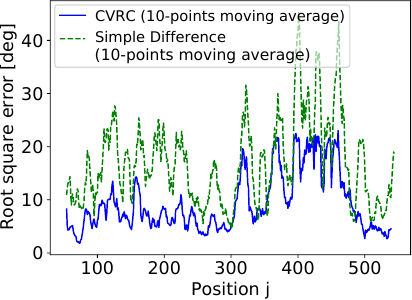}
    \subcaption*{(b-2)}
  \end{minipage}
  \caption{($\star$-1) Estimation results of CVRC, neighbor difference method, and the ground truth and ($\star$-2) root square errors of these methods in the data (a-$\star$) used for learning and (b-$\star$) unused for learning.}
  \label{fig:est-slope}
\end{figure}

  Fig.~\ref{fig:est-slope} shows ($\star$-1) the estimation results of CVRC, neighbor difference method, and the ground truth and ($\star$-2) root square errors of these methods with (a-$\star$) the data used for learning {($i=300$)} and (b-$\star$) the data unused for learning {($i=500$)}.
In the area used for learning, the CVRC estimation error is about 4.8 deg while the error in the neighbor difference result is 12.4 deg in average.
The CVRC result is robust against noise showing a higher accuracy than the neighbor difference method in the area unused for learning.

\section{Conclusion}
\label{s:conclusion}

We have proposed CVRC to deal with complex-valued data of interferogram obtained in InSAR.
We fed the two-dimensional spatial data to the CVRC to classify local land forms.
As a result, we found that CVRC achieves higher accuracy and more strength against noise than RVRC.
CVRC performs classification successfully without resolution degradation and high computational cost unlike CVCNN.

In addition, we presented that CVRC has a high adaptability to process complex-amplitude data through our experiments of the aspect classification and the slope angle estimation.
In the near future, CVRC will play an important role to deal with large amount of InSAR data by utilizing its high speed processing and generalization ability.

\section*{Acknowledgement}
The authors thank Gouhei Tanaka, Project Associate Professor, International Research Center for Neurointelligence (IRCN), the University of Tokyo, and Ryosho Nakane, Project Associate Professor, Department of Electrical Engineering and Information Systems, the University of Tokyo, for their advice on experimental design.

\bibliography{sar,sar_app,nn,rc,rc_app}

\end{document}